\documentclass[a4paper,twocolumn,11pt]{quantumarticle}
\pdfoutput=1
\usepackage[utf8]{inputenc}
\usepackage[english]{babel}
\usepackage[T1]{fontenc}
\usepackage{amsmath}
\usepackage[colorlinks,breaklinks,linkcolor=blue,anchorcolor=blue,citecolor=blue,urlcolor=blue]{hyperref}
\usepackage{subfigure}
\usepackage{tikz}
\usepackage{lipsum}
\usepackage{graphicx}
\usepackage{epstopdf}

\begin{document}

\title{Quantum indirect synchronization}
\author{GuoMeitong}
\affiliation{Center for Quantum Sciences and School of Physics, Northeast Normal University, Changchun 130024, China}
\author{NiuXiangyu}
\orcid{0000-0003-0290-4698}
\affiliation{Center for Quantum Sciences and School of Physics, Northeast Normal University, Changchun 130024, China}
\author{YiXuexi}
\affiliation{Center for Quantum Sciences and School of Physics, Northeast Normal University, Changchun 130024, China}
\affiliation{Center for Advanced Optoelectronic Functional Materials Research, and Key Laboratory for UV-Emitting Materials and Technology of
Ministry of Education, Northeast Normal University, Changchun 130024, China}
\author{WangWei}
\affiliation{School of Physics, Northeast Normal University, Changchun 130024, China}
\email{wangw100@nenu.edu.cn}
\maketitle

\begin{abstract}
  It is well known that a system with two or more levels exists a limit cycle and can be synchronized with an external drive when the system and the drive are directly coupled. One might wonder if a system can synchronize with the external drive when they are not coupled directly. In this paper, we examine this case by considering a composite system consisting of two coupled two-level quantum systems, one of which is driven by an external field, while another couples to the driven one. Due to the decoherence caused by environments, the composite system would stay in a mixed state, and an effective limit cycle is formed, so phase locking could occur. We find the phase locking phenomenon in the phase diagram characterized by Husimi $Q$ function, and the synchronization can be generated consequently that we will refer to indirect synchronization. The $S$ function defined in the earlier study can also be used to measure the strength of synchronization. We claim that indirect synchronization is possible. This result provides us with a method to synchronize a quantum system that coupled to its neighbour without interacting with external drive directly.
\end{abstract}

As a collective dynamic feature of complex systems, synchronization is an ubiquitous phenomenon in physics. The study of synchronization can be dated back to the 17th century when Huygens discovered an odd kind of sympathy between two pendulum clocks, their oscillations show synchronization phenomenon when coupled to each other via a common support \cite{H. Fujisaka and T. Yamada(1983), L. M. Pecora and T. L. Carroll(1990), M. Barahona and L. M. Pecora(2002), C Huygens}. Huygens then noticed that the oscillations of these pendulum clocks tended to synchronize with each other. Since then, synchronization has been extensively studied and widely applied in classical physics, where researchers find that the phases between the phase locking oscillator and the external signal spontaneously synchronize in their respective phase space trajectories.

The extension of the synchronization concept from classical to quantum physics has been suggested \cite{I. Goychuk,M. Ludwig,A. Mari,T. E. Lee,A. M. Hriscu,G. Manzano,S. E. Nigg}, and widely been applied in the fields of cavity quantum electrodynamics \cite{O. V. Zhirov and D. L. Shepelyansky(2009),V. Ameri(2015)}, masers \cite{C. Davis}, atomic combination \cite{M. Xu(2014), M. Xu and M. J(2015),M. R. Hush(2015)}, Kerr-anharmonic \cite{N. Lorch(2017)}, Van der Pol (VDP) oscillator \cite{ T. E. Lee and H. R(2013), T. E. Lee(2014), S. Walter(2014)}, Bose-Einstein condensate \cite{M. Samoylova(2015)}, and superconducting circuit system \cite{Y. Gul(2016),Fernando}. Though most works focus on the research of the quantum synchronization of position and momentum, studies on synchronization of the other degrees of freedom become active recently. Roulet and Bruder \cite{Alexandre Roulet(1)} have recently proved that three or more levels quantum system in pure state can synchronize with the external field, which possesses all the synchronization features that classical system has. The authors further pointed out that a two-level system in pure states is impossible to synchronize because the system has no stable limit cycle. But it can be synchronized in mixed states as \cite{Alvaro Parra-Lopez(2020),E. H} shown. They proved the synchronization by solving the Lindblad master equation of the two-level system to simulate self-sustained mixed state. The result shows that it supports a valid limit cycle, and phase locking \cite{Alexandre Roulet(1)} can be found in this system different from that in classical deterministic systems.  They also calculated the synchronization measurement that characterizes the phase locking strength and analyzes the dependence of the signal strength on the natural frequency of the two-level system. Except the finite level systems, synchronization can also be observed, for example, in a two-node network consisting of the self-contained oscillator of spin-1. It is shown that even if the situation that limit cycle can not synchronize with external semi-classical signals, phase locking still can be established, which may help setting quantum node networks \cite{Alexandre Roulet(2)}.

Regardless of these progresses made in quantum synchronization, the study of synchronization between a drive and its indirectly driven object (we will name as indirect synchronization throughout this paper) is in its infancy so far. Inspiring by the control theory developed in recent years \cite{G. M,R. S,H. M,A. C1,A. C2,A. C3,P. H,S. Lloyd, D. Burgarth,A. Kay,M. Owari,Kato}, we naturally ask if it is possible to indirectly synchronize a quantum system with a drive. At first sight, this question is trivial, however, extensive examination shows that the features of indirect synchronization are far beyond speculation. Besides, drives may be difficult to directly applied to a system in practice, so adding a drive to the other system coupling strongly to the system of interests to achieve the synchronization might find practical applications.

The paper is organized as follows. In Sec. \ref{Sec:Model}, we introduce our model of two coupled quantum two-level systems where one subsystem is driven by an external field. In Sec. \ref{Sec:Qfunction}, we investigate the synchronization between the drive and the subsystem indirectly coupled to the drive, and discuss the limit cycle of the system. By calculating and examining the Husimi $Q$ function, the features of the synchronization are analyzed. In Sec. \ref{Sec:Synchronization}, we quantify the degree of phase locking and the synchronization. The paper finally concludes in Sec. \ref{Sec:Conclusion} with a summary of our main results.

\section{Model}\label{Sec:Model}
\begin{figure}[htbp]
  \centering
  \includegraphics[width=8cm]{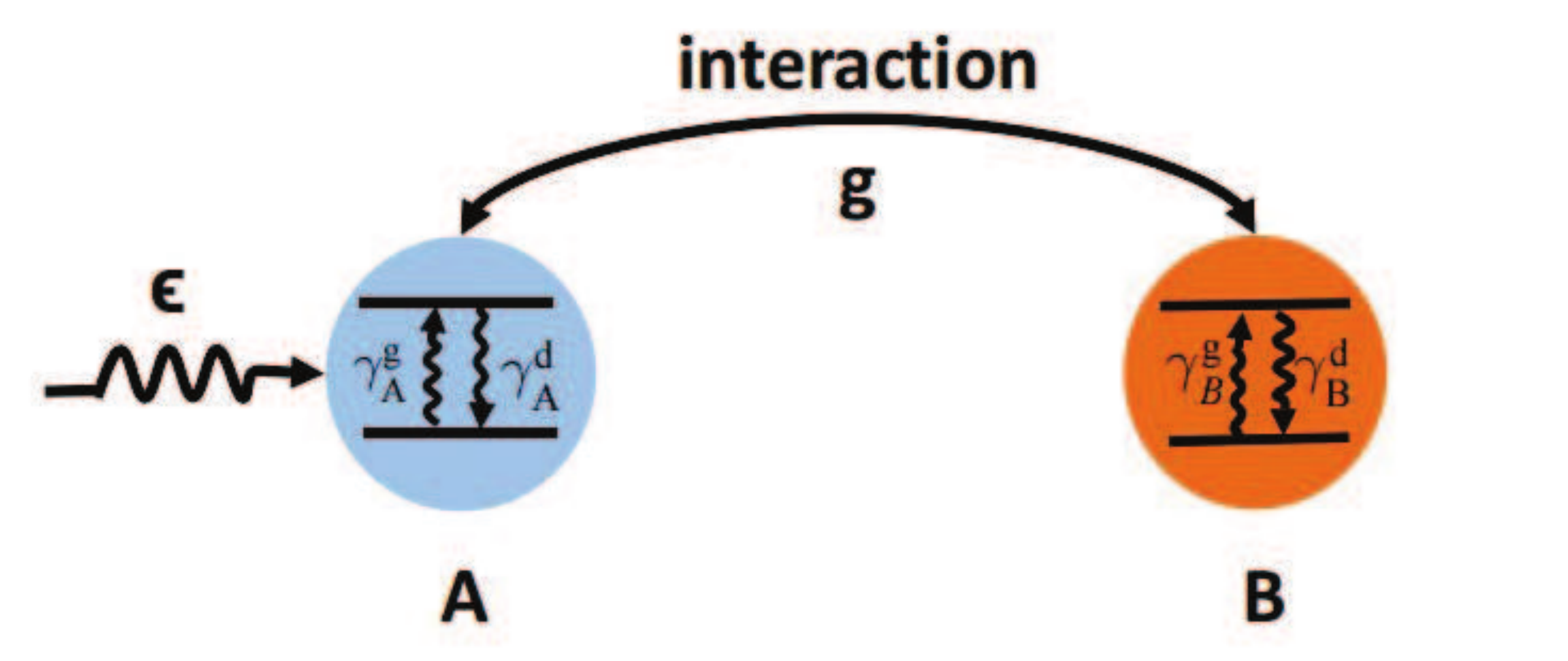}
  \caption{Illustration of our system. Two two-level subsystems $A$ and $B$ are coupled with coupling constant $g$. The drive is directly exerted on the subsystem $A$ and indirectly coupled to the subsystem $B$ via $A$. Both subsystem $A$ and $B$ are coupled to the environments separately, and we denote $\gamma _j^g$ as the gain rate, while $\gamma _j^d$ as the loss  rate ($j=A,B$).}
  \label{2DBS}
\end{figure}

We consider coupled two-level systems $A$ and $B$ with natural frequency $\omega _A$ and $\omega _B$, respectively, as shown  in Fig. \ref{2DBS}. For such a system, the free Hamiltonians read

\begin{equation}
{\hat H}_j = \frac{1}{2}\hbar {\omega _j}\hat{\sigma}_j^{z},
\end{equation}
where $j=A, B$, and $\hat{\sigma}_j^{z}$ are Pauli matrices.

The interaction between the subsystem $A$ and the subsystem $B$ can be described by
\begin{equation}
{{\hat V}} = i\hbar \frac{{{g}}}{2}(\hat \sigma _A^ + \hat \sigma _B^- - \hat \sigma _B^+ \hat \sigma^- _A ) ,
\end{equation}
where $g$ is the coupling strength between the subsystem $A$ and the subsystem $B$, $\hat \sigma _{\rm{j}}^ + (\hat \sigma _j^ - )$ is the raising (lowering) operator of spin $j$ $(j=A,B)$ and ${\hat \sigma_j^\pm }=\frac{1}{2}\left( {{{\hat \sigma }_j^{x}} \pm i{{\hat\sigma}}_j^{y}} \right)$.
We use a classical field with frequency $\omega$ and strength $\epsilon$. In the rotating-wave approximation \cite{Michel Le Bellac(2006)}, the Hamiltonian takes
\begin{equation}
{\hat{H}_{{\rm{signal}}}} = i\hbar \frac{\epsilon}{2}({e^{i\omega t}}\hat \sigma_{A}^-- {e^{ - i\omega t}}\hat \sigma _{A}^+ ) .
\end{equation}
In the frame rotating with $\hbar\omega\hat{\sigma}_z$, the dynamics of the system ($A$ plus $B$) is governed by the master equation
\begin{equation}
\begin{split}
\dot{\hat{\rho}}_{AB}=&-{i}[\hat{V},\hat{\rho}_{AB}]  { - \frac{i}{2}[{\Delta _1}\hat \sigma _{A}^z+{\Delta_2}\hat\sigma_{B}^z{\rm{ + }}\epsilon \hat \sigma _{A}^y,\hat \rho_{AB} ]}\\  &+ \sum\limits_{j = A,B} {\mathcal{L}_j}\hat \rho_{AB} ,
\end{split}
\end{equation}
here we set ${\hbar=1}$, ${\hat\rho}_{AB}$ is the density matrix of the system, $\Delta_{1,2}= \omega _{A,B} - \omega$ is the detuning of the external drive frequency to the natural frequency of the subsystem $A,B$. Lindblad dissipator superoperator ${\mathcal{L}_j}\hat \rho  = \frac{{\gamma _j^g}}{2}D[\hat \sigma _j^ + \hat \sigma _j^z]\hat \rho  + \frac{{\gamma _j^d}}{2}D[\hat \sigma _j^ - \hat \sigma _j^z]\hat \rho $ and the Markovian master equation is written in the standard Lindblad form \cite{H.P. Breuer} as $D[\hat O]\hat \rho  = \hat O\hat \rho {\hat O^\dagger } - \frac{1}{2}\{ {\hat O^ \dagger }\hat O,\hat \rho \} $, and $\gamma_j^g$ is the gain rate, $\gamma_j^d$ is the loss rate.

\section{Husimi $Q$ function}\label{Sec:Qfunction}

By solving the master equation, we can obtain the steady state density matrix $\hat\rho^S _{AB}$ of the system $A$ plus $B$,
\begin{equation}
{\hat\rho^S _{AB}} = \sum_{\mathord{\buildrel{\lower3pt\hbox{$\scriptscriptstyle\rightharpoonup$}}
\over m} ,\mathord{\buildrel{\lower3pt\hbox{$\scriptscriptstyle\rightharpoonup$}}
\over n} } {{\hat \rho }_{{\rm{(}}\mathord{\buildrel{\lower3pt\hbox{$\scriptscriptstyle\rightharpoonup$}}
\over m} ,\mathord{\buildrel{\lower3pt\hbox{$\scriptscriptstyle\rightharpoonup$}}
\over n} {)}}\vert{m_A},\left. {{m_B}} \right\rangle \left\langle {{n_A},\left. {{n_B}} \right\vert} \right.},
\end{equation}
where $\vec m = (m_A,m_B)$, $\vec n = (n_A,n_B)$, and $m_{A(B)}$ and $n_{A(B)}$ denote the quantum numbers of subsystems $A$ and $B$. For the systems under consideration,  $m_{A(B)}=n_{A(B)}=\pm1/2$.

In order to evaluate the synchronization of the subsystem $B$ with the external signal, we perform a partial trace over the subsystem $A$ to obtain the reduce density operatr of $B$,  $\hat{\rho}_B = \text{Tr}_A[\hat{\rho}^S_{AB}]$. The density matrix $\hat{\rho}^S_{AB}$ is the steady-state density matrix $\hat{\rho} _{B}$ for spin $B$.
To visualize the behavior of the system, we use the Husimi $Q$ representation ($Q$ function) as a measure to quantify the phase of the spin system \cite{R. Gilmore(2018)}. It is a quasi-probability distribution that allows us to represent the two-level system in the phase space. The $Q$ function is defined by \cite{Alexandre Roulet(1)}
\begin{equation}
Q(\theta,\phi)=\frac{1}{2\pi}\langle\theta,\phi \vert{\hat{\rho}_{B}}\vert\theta,\phi\rangle ,
\end{equation}
This spherical representation is formulated in terms of spin-coherent states $\vert\theta ,\phi \rangle$ \cite{Arecchi F T}, which in the case of a two-level system (TLS) are the eigenstates of the spin operator $\sigma_{\mathbf{n}} = \mathbf{n} \cdot \boldsymbol{\hat{\sigma}}$, i.e., the spin operator along the direction of the unit vector $\mathbf{n}$. Here $\theta$ and $\phi$ are the polar coordinates. Examining $Q$ function, we can find the synchronization features of the system under study.
\begin{figure}[htbp]
  \centering
  \includegraphics[width=7cm]{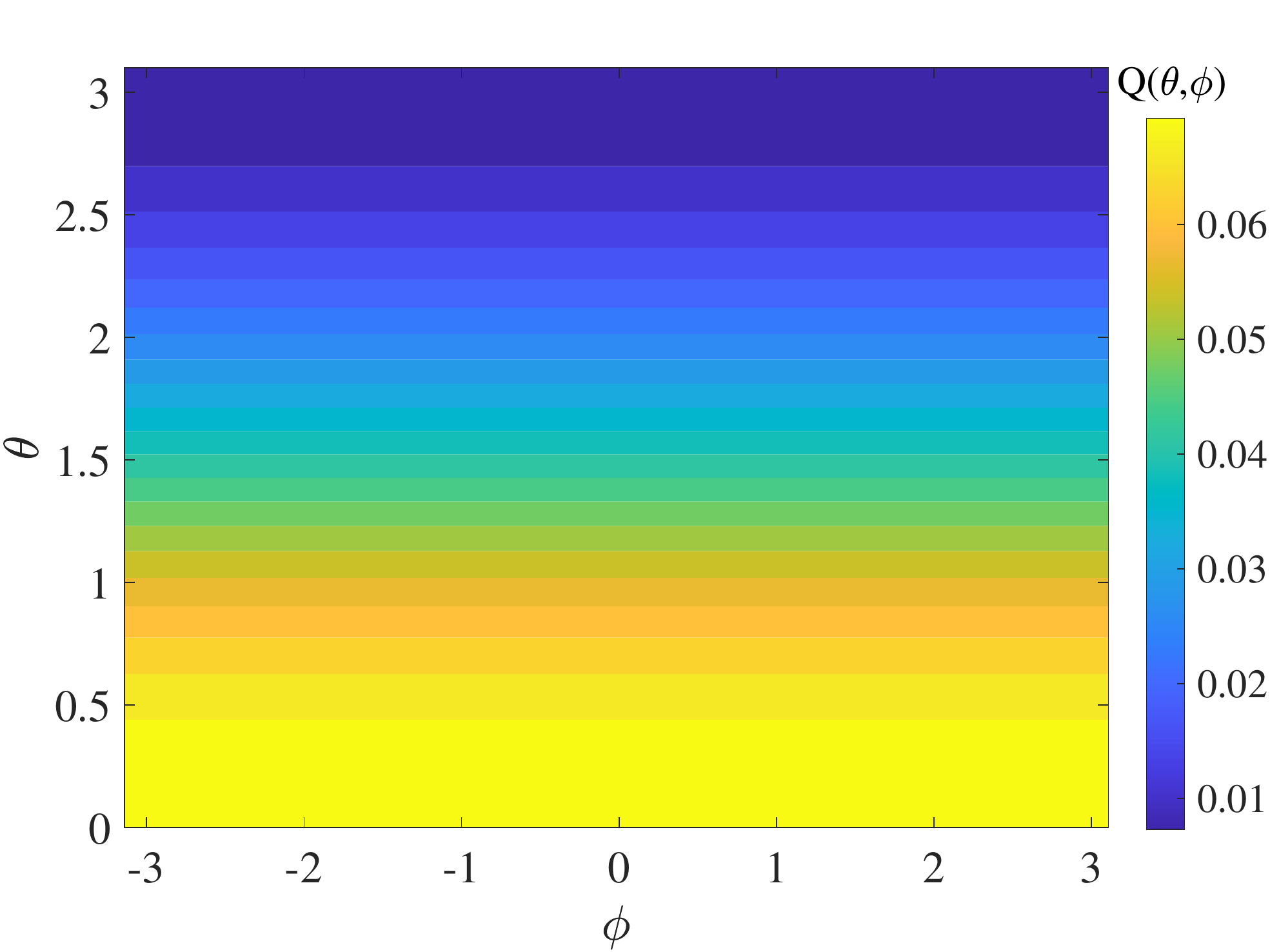}
  \caption{The limit cycle. When there is no signal, the subsystem $B$ is isolated. We observe that the distribution is independent of the $\phi$, and the distribution is centered on the $\theta = 0$, corresponding to the coherent state describing the spin precessing around  the $z$-axis.}
  \label{QFg}
\end{figure}

\begin{figure}[htbp]
\centering
  \subfigure{
  \label{QFd}
  \includegraphics[width=4.75cm,height=3.75cm]{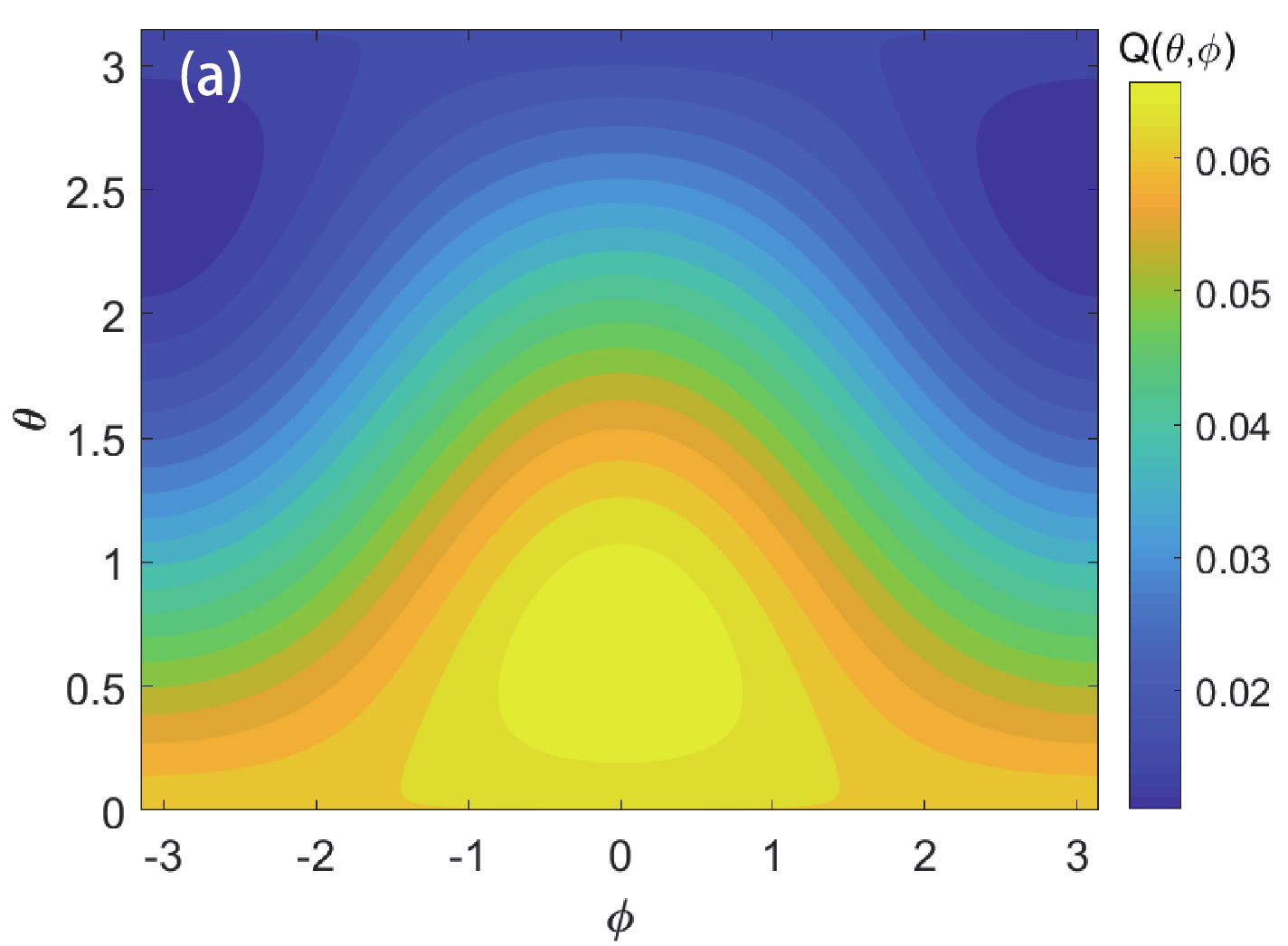}}\subfigure{\label{QFe}
  \includegraphics[width=4.75cm,height=3.75cm]{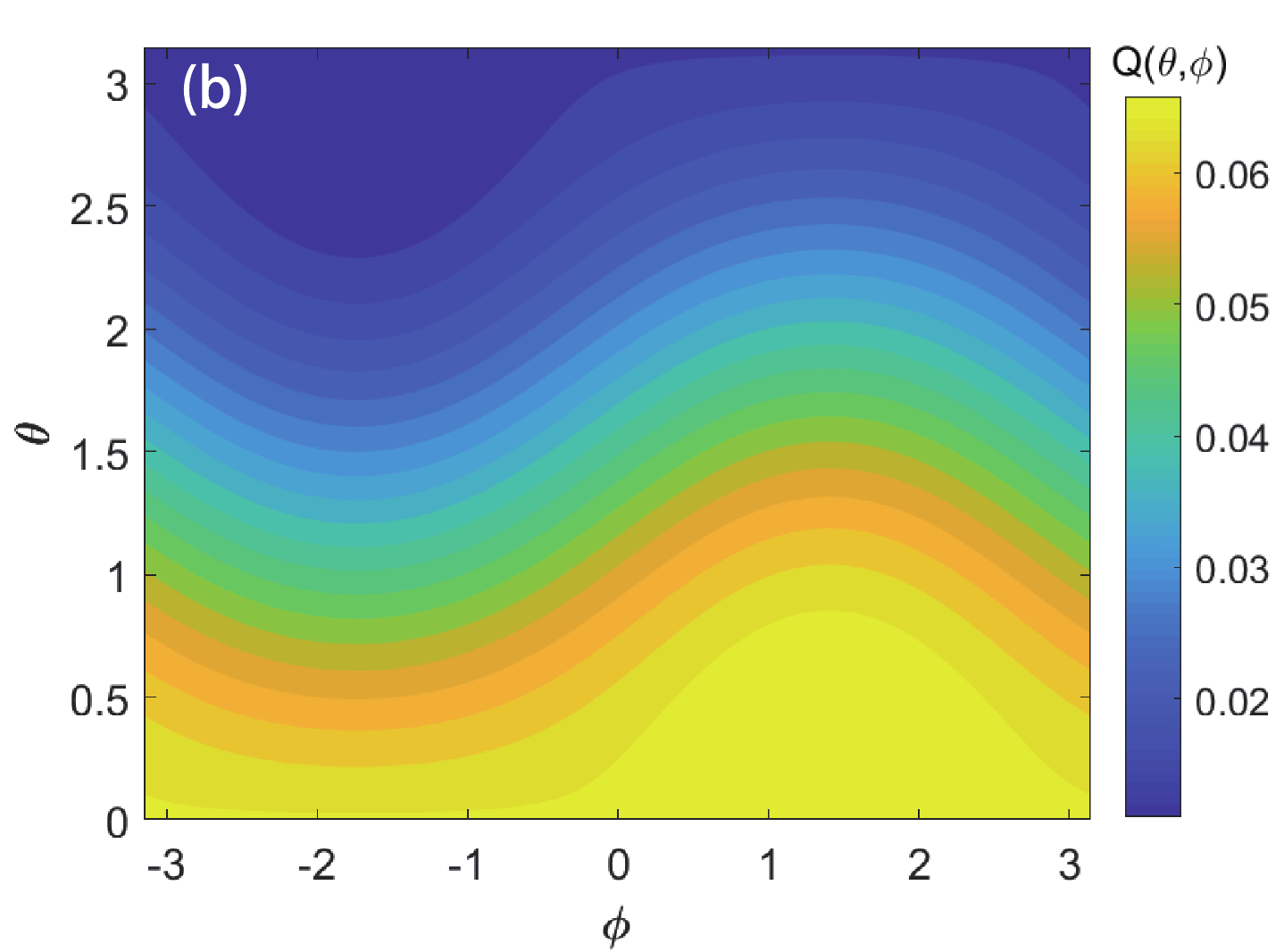}}
  \caption{(a) The steady-state Q function depending on $\theta$ and $\phi$. The steady-state Q function represents phase locking to a resonant external signal. ${\Delta _1} = {\Delta_2} = 0$, $\epsilon = 5\gamma _B^d$, $g=8\gamma _B^d$, $\gamma _A^g/\gamma _A^d = 10$, $\gamma _B^g/\gamma _B^d = 10$ and $\gamma _A^g = \gamma _B^g$. (b) The steady-state Q function depending on $\theta$ and $\phi$. The steady-state Q function represents phase locking to a non-resonant external signal. ${\Delta _1} = 3$, ${\Delta _2} = 5$, $\epsilon  = 5\gamma _B^d$, $g = 8\gamma _B^d$, $\gamma _A^g/\gamma _A^d = 10$, $\gamma _B^g/\gamma _B^d = 10$ and $\gamma _A^g = \gamma _B^g$.}

\end{figure}

When the coupled strength $g=0$, the Husimi $Q$ of subsystem $B$ is,  $Q(\theta ,\phi ) = (11 + 9\cos{(\phi)})/{88\pi}$, which is shown in Fig. \ref{QFg}. It is independent of $\phi$ and centered around $\theta=0$, corresponding to coherent states on the equator precessing about the $z-$axis. The signal not only attracts the phases towards to $\theta  = 0$, but also makes the system having no $\phi$ phase preference. It is because that the master equation is invariant by rotating operation around the z-axis. Because the state  has no a definite $\phi$ preferred  region, the system is not of phase locking. Then synchronization feature can not be displayed when the $g=0$ similar to the results when there is no signal ($\epsilon$)  as previous research shows \cite{Alvaro Parra-Lopez(2020)}.

For the steady-state $Q$ function with $g\ne 0,\epsilon\ne 0$ and  no detuning between the drive and subsystem $B$, the results are shown in  Fig. \ref{QFd}. We find from the figure that a peak in the $Q$ function appears around $\theta {=}0$ and $\phi{=}0$, indicating that system is phase locking.
With non-zero detuning between the drive and subsystem $B$, we find from Fig. \ref{QFe} that  the peak of the $Q$ function  appears near 0.  The detuning shifts the phase locking towards $\phi=\frac{\pi}{2}$. In the next section,  we attempt to measure how strong the synchronization is via a synchronization measure.

\section{Synchronization}\label{Sec:Synchronization}
To quantify the synchronization, we will use  the $Q$ function \cite{Alexandre Roulet(1)} as a tool to measure the synchronization. In the last section we find that when the  drive $\epsilon=0$ or the strength of coupling $g=0$, the steady-state $Q$ function distributes on both sides of $\theta=0$ and since the dissipative source of energy does not favor any phase $\phi$, there is no phase locking for the subsystem $B$. When the drive is applied  to the subsystem $A$,  phase locking in subsystem $B$ occurs. From the phase space Husimi $Q$ representation, we can obtain the phase distribution $S(\phi)$ of the state ${\hat\rho_B}$ by integrating out the $\theta$ angle. Since the dissipative source of energy does not have any phase preference of the subsystem $B$, the noise leads to phase diffusion, and which leads to the limit-cycle state $\hat\rho_B$ with a united phase distribution $S(\phi)= 1/4\pi$. To quantify the strength of the phase locking or the synchronization for the subsystem $B$, we define a measure $S$ as
\begin{equation}
S(\phi) = \int _0 ^{\pi} d\theta \sin{\theta}Q(\theta, \phi) - \frac{1}{4\pi}.
\end{equation}
$S(\phi)$ can be seen as a peak height of the phase locking above a flat distribution. It is zero when there is no synchronization. The uneven distribution of the phase will result in a non-zero value of the measure, indicating a locking on the corresponding phase \cite{N. Lorch(2017), M. R. Hush(2015)}.

\begin{figure}[h]
  \centering
  \includegraphics[width=7cm,height=5cm]{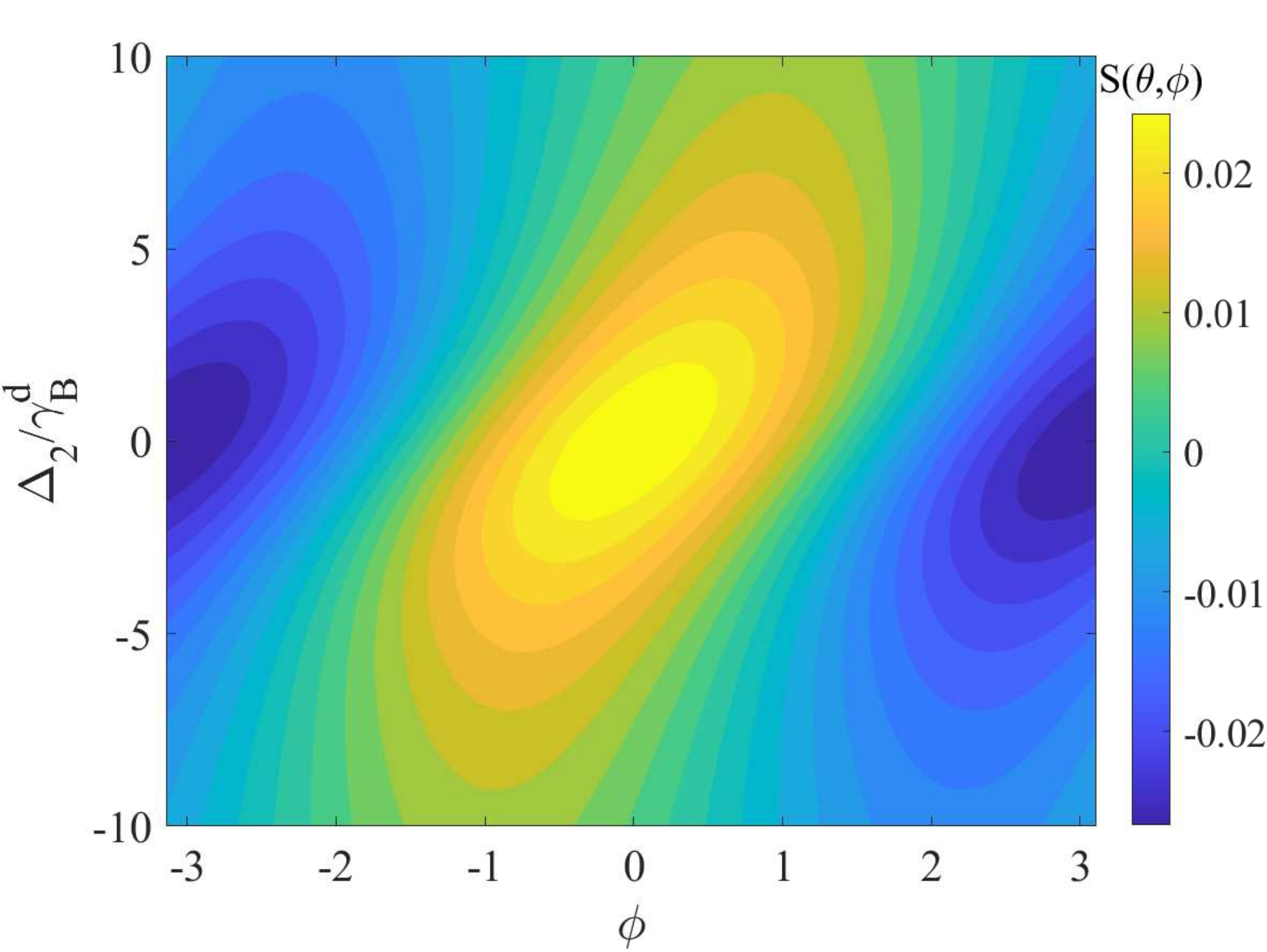}
  \caption{The distribution of $S(\phi)$ function varies with phase $\phi$ and detuning ${\Delta_2}$. $\epsilon  = 5\gamma _B^d$, ${\Delta _1} = 0$, $g = 8\gamma _B^d$, $\gamma _A^g/\gamma _A^d = 10$, $\gamma _B^g/\gamma _B^d = 10$ and $\gamma _A^g = \gamma _B^g$. When there is no detuning, the phase locking is stronger. When $\Delta_2$ is positive or negative, the maximal value of $S(\phi)$  shift  towards $\phi = \pi$ or $\phi = -\pi$, severally.}
	\label{ATa}
\end{figure}

We look for a signature of synchronization by calculating the phase distribution from the steady-state solution of the master equation. In Fig. \ref{ATa}, we show $S(\phi )$ distribution as a measure of phase locking. The subsystem $B$ shows signatures of synchronization as $S(\phi)$ is not flat. As we take different values of detuning $\Delta_2$, the phase locks at different values. The higher value of $S(\phi)$ is around $\Delta_2=0$ and $\phi=0$.

\begin{figure}[h]
  \centering
  \includegraphics[width=7cm,height=5cm]{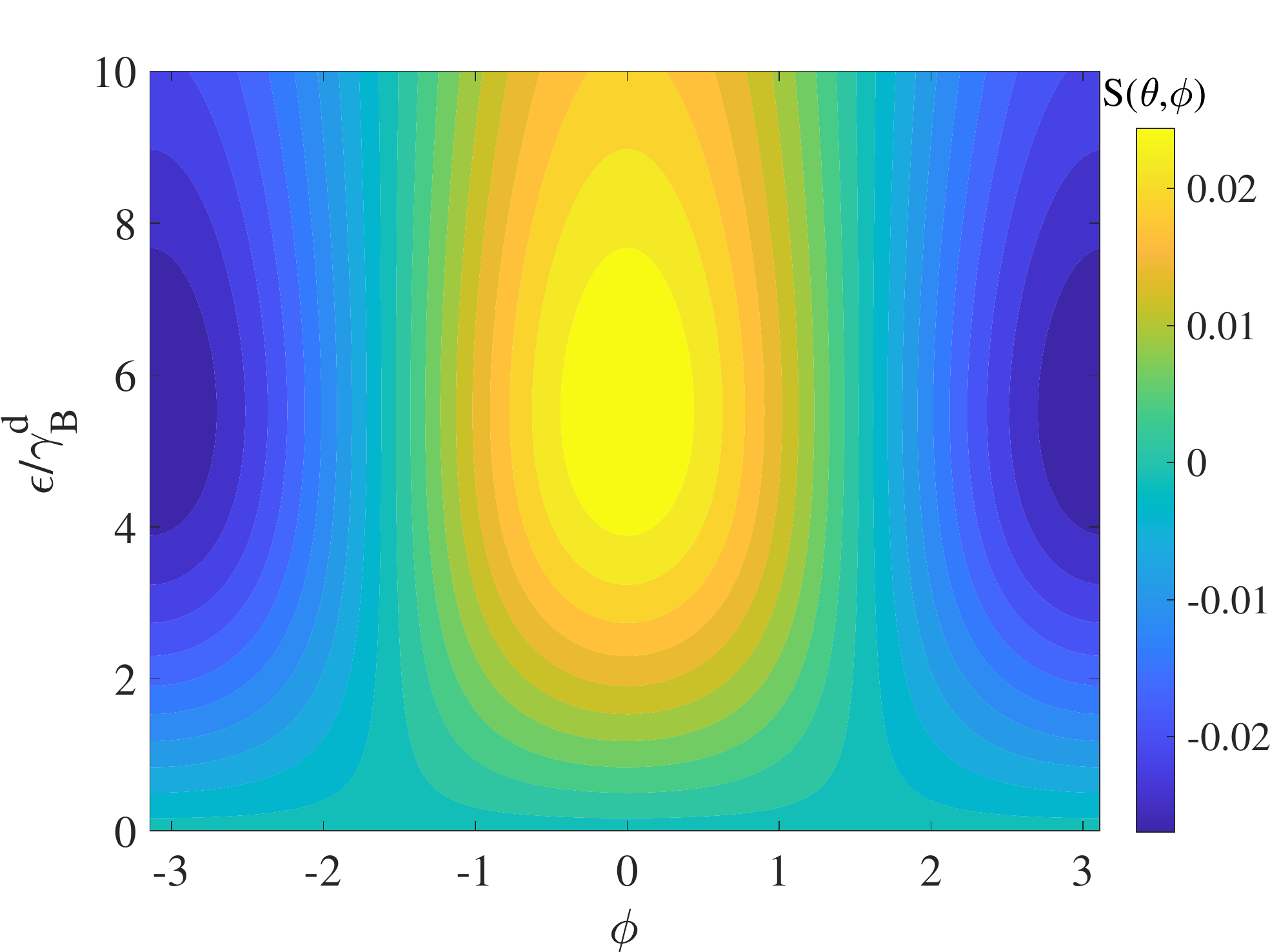}
  \caption{The distribution of $S(\phi)$ function varies with phase $\phi$ and the strength of the signal $\epsilon$. $\Delta_2 = 0$, ${\Delta _1} = 0$, $g = 8\gamma_B^d$, $\gamma_A^g/\gamma _A^d = 10$, $\gamma_B^g/\gamma_B^d = 10$ and $\gamma_A^g =\gamma_B^g$. When $g$ is fixed, we investigated the effect of signal intensity on phase locking intensity. As the strength of the signal increases, it is observed that the phase locking is stronger. It is because there is no detuning, the highest values locate at $\phi = 0$, as we could infer from Fig. \ref{ATa}.}
	\label{ATb}
\end{figure}

The relationship between signal strength and synchronization measures is shown in Fig. \ref{ATb}. When $g$ is fixed, we find that $S(\phi)$ tends to be more and more localized as we increase the signal strength $\epsilon$. The synchronization feature is displayed. The synchronization is stronger with a greater value for $\epsilon$, locating at $\phi = 0$. The phenomenon is just like that synchronization in TLS, but a larger $\epsilon$ is needed to achieve the same strength of synchronization. We find that when the signal strength reaches a certain value, the synchronization effect is the best. When the signal strength continues to increase, the synchronization effect becomes weaker and weaker. A excessive strong signal will take the system out of synchronization \cite{Koppenhofer}, by definition, the synchronization can be realized only if the strength of the signal $\epsilon$ is small enough so that the limit cycle is only slightly disturbed. When the strength of the signal beyond this range means not only affects the phase of the system oscillation, but also affects its amplitude. Thus, the limit cycle is deformed, so the strength of the signal should be in a proper range that can be synchronized.

\begin{figure}[h]
  \centering
  \includegraphics[width=7cm,height=5cm]{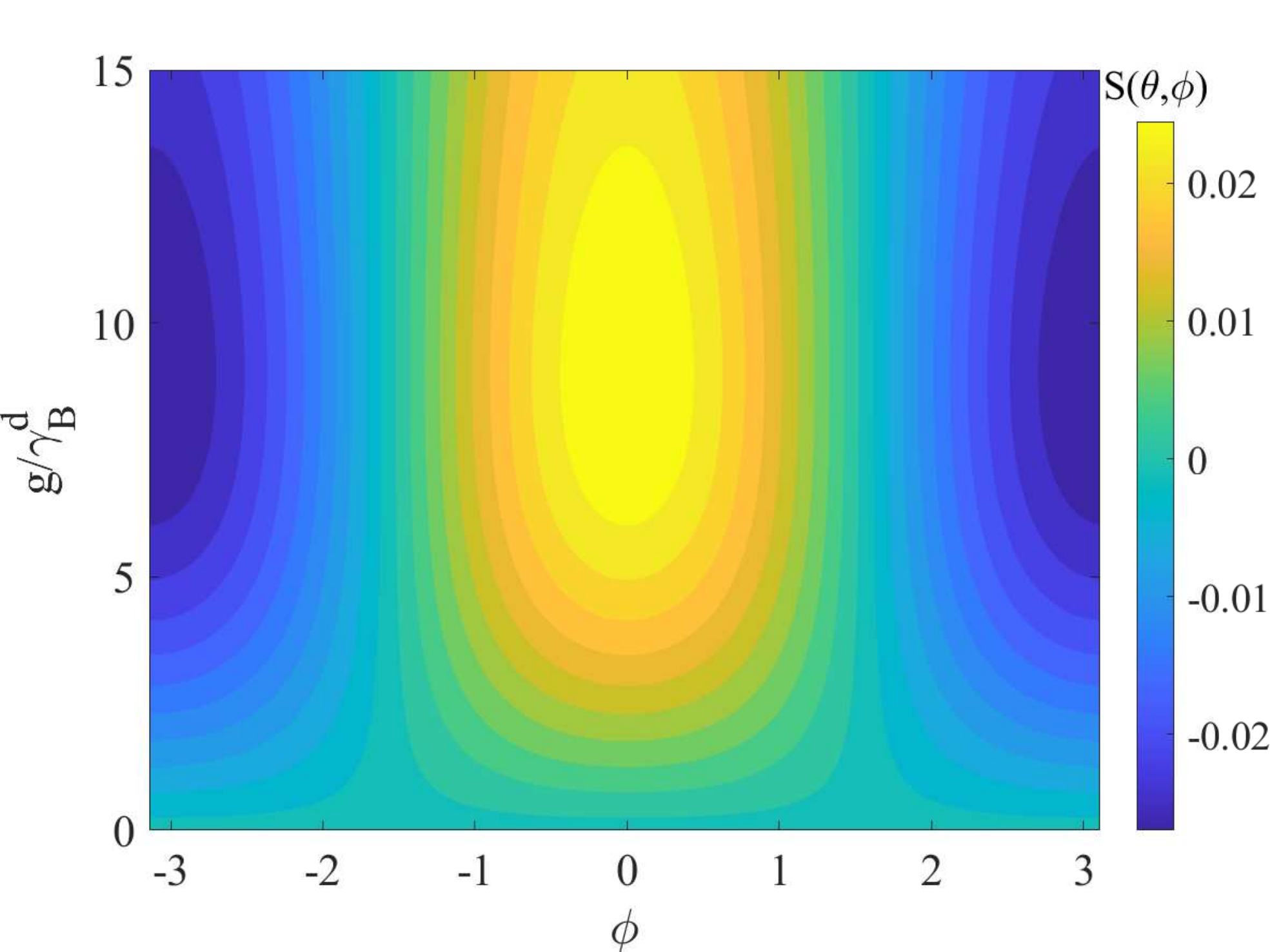}
  \caption{The distribution image of $S(\phi)$ function varies with phase $\phi$ and the strength of coupling $g$. $\Delta_2 = 0$, ${\Delta _1} = 0$, $ \epsilon= 5\gamma _B^d$, $\gamma _A^g/\gamma _A ^d = 10$, $\gamma _B^g/\gamma _B^d = 10$ and $\gamma _A^g = \gamma _B ^g$. When $\epsilon$ is fixed, we study how the strength of coupling modifies the strength of the phase locking. We find that the greater the strength of coupling, the more obvious phase locking phenomenon is. Because there is no detuning, the highest values locate at $\phi = 0$.}
	\label{ATc}
\end{figure}

Since in our system the drive is subjected to the subsystem $A$, and there is a coupling between subsystems $A$ and $B$, therefore we consider the coupling strength between two systems. In Fig. \ref{ATc}, we consider the effect of the coupling strength on synchronization. It shows that when $\epsilon$ is fixed, increasing the coupling strength $g$ results in more distinct phase locking. Thus, we know that in the control of indirect synchronization, when the driving strength is constant, the system synchronization effect with higher coupling strength is more obvious. We find that when the coupling strength reaches a certain value, the synchronization effect is the best. As the signal strength continues to increase, similarly the synchronization effect goes down.

\begin{figure}[h]
  \centering
  \includegraphics[width=7cm,height=5cm]{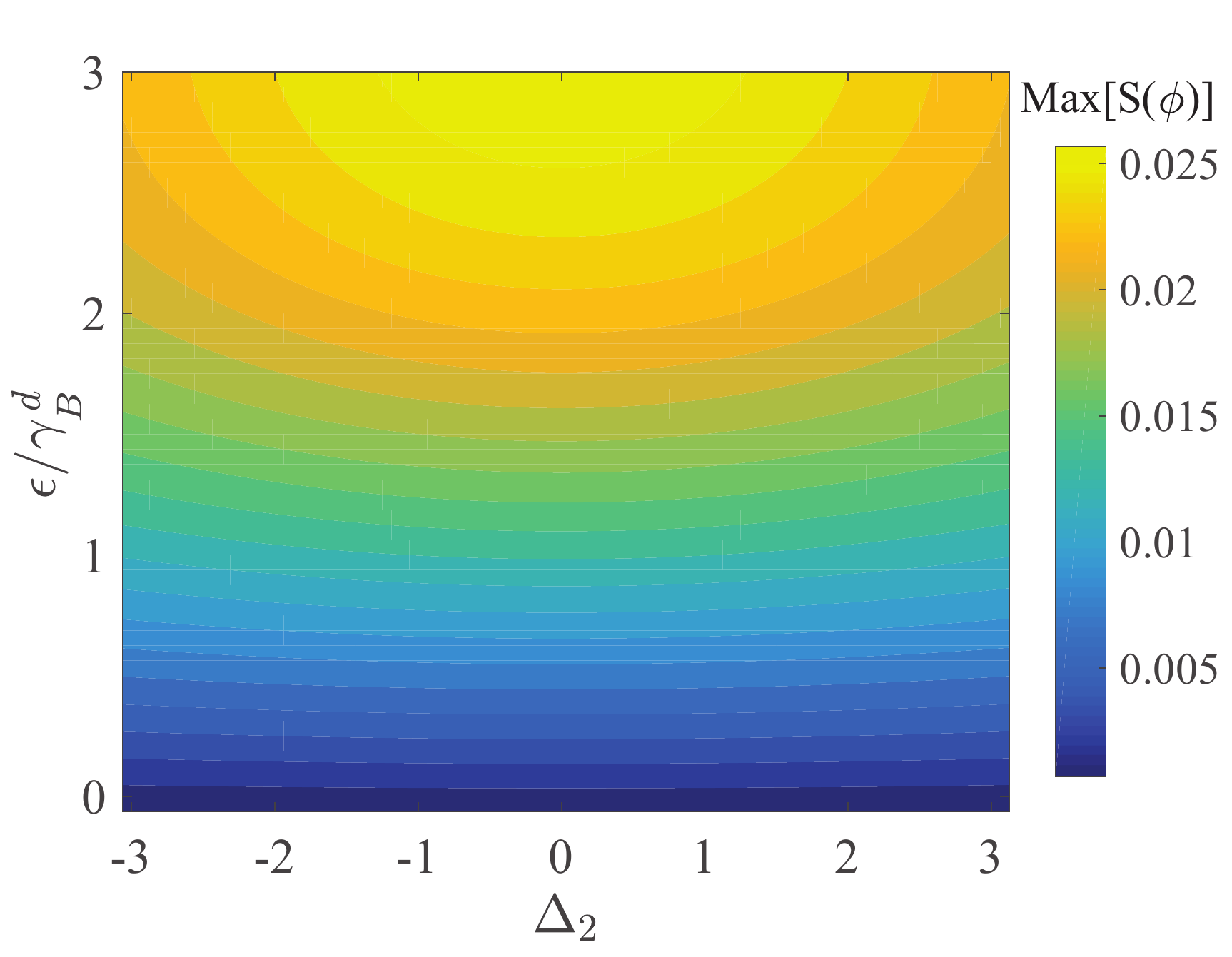}
  \caption{Arnold tongue of the system. We plot the maximum value of $S(\phi)$ as function of the strength $\epsilon$ and the detuning $\Delta_2$, with $g = 8\gamma _B^d$, $\gamma _A^g/\gamma _A^d = 10$, $\gamma _B^g/\gamma _B^d = 10$ and $\gamma _A^g = \gamma _B^g$.}
	\label{ATz}
\end{figure}

\begin{figure}[h]
  \centering
  \includegraphics[width=7cm,height=5cm]{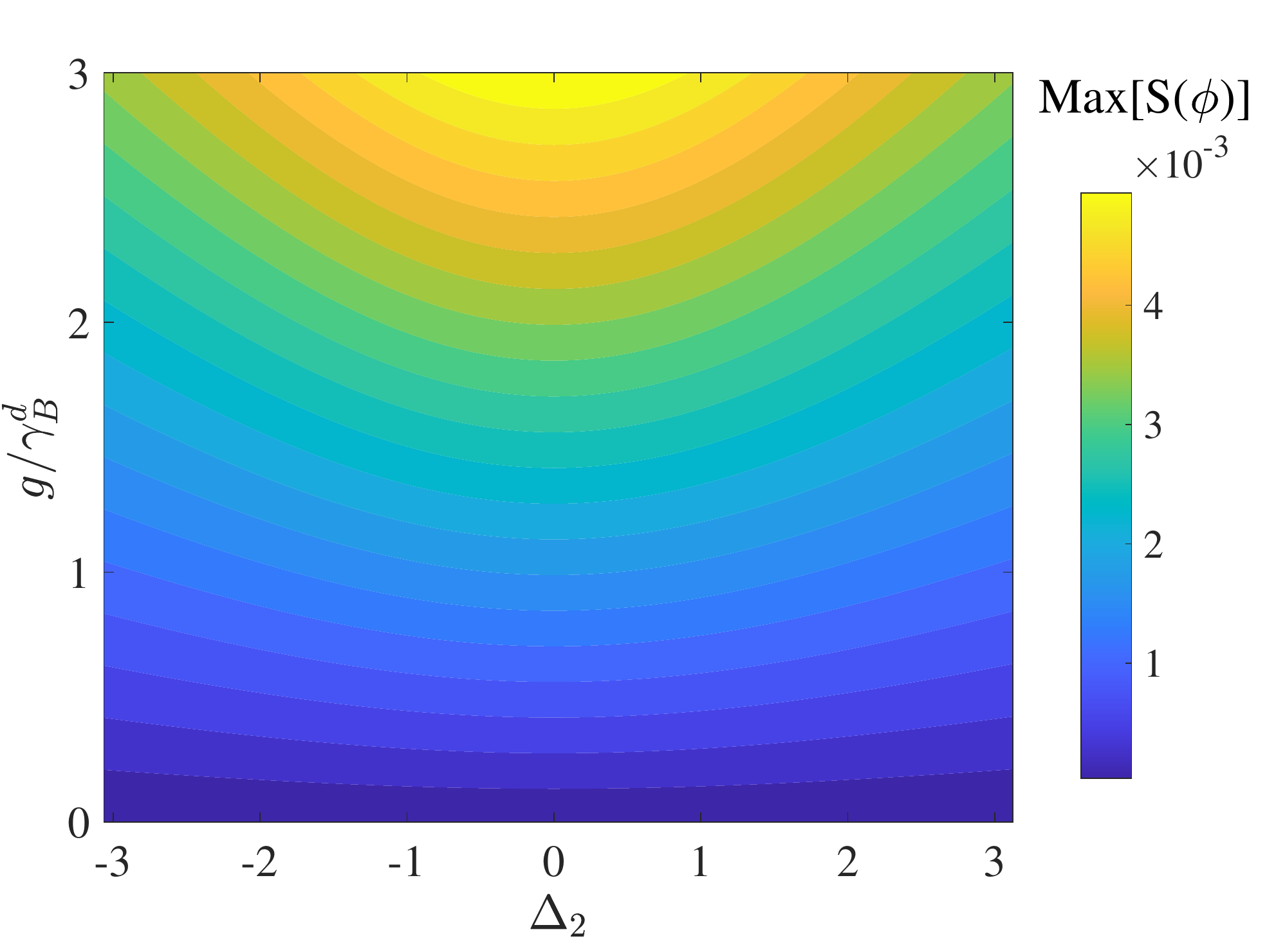}
  \caption{Arnold tongue of the system. We plot the maximum value of $S(\phi)$ as function of the coupling strength $g$ and the detuning $\Delta_2$, with $\epsilon = 5\gamma _B^d$, $\gamma _A^g/\gamma _A^d = 10$, $\gamma _B^g/\gamma _B^d = 10$ and $\gamma _A^g = \gamma _B^g$. If $\epsilon = 0$ or $g = 0$, $Max[S(\phi)]$ will be zero. Otherwise $Max[S(\phi)]$ will always give non-zero, even if it is so small that the synchronization can be ignored.}
	\label{ATd}
\end{figure}

We consider the maximum of $S(\phi)$ as a measure of synchronization. Fig. \ref{ATz} relating to a set of the signal strength $\epsilon$ and detuning values that occur phase locking. The stronger the signal strength $\epsilon$, the larger the range of detuning $\Delta_2$ leading to significant localization of the relative phase, also yielding Arnold tongue \cite{S. Sonar} shown in Fig. \ref{ATd}. Due to quantum noise, a small $\epsilon$ nearly shows a vanishing strength of synchronization, thus at this point shall be qualitatively chosen to ensure that for any detuning value, it will not distort the limit cycle. \cite{H. J. Carmichael}.

Finally, Fig. \ref{ATd} relates to a set of coupling strengths and detuning values that occur phase locking. To a certain extent, when the detuning is very small and the coupling strength is large enough, it is easy to synchronize. These similar results can be seen in the previous researched direct synchronization scheme \cite{Alexandre Roulet(1)}, and the coupling strength in this system is just like the drive strength of direct synchronization, leading the subsystem B into the region of synchronization, which also imply the validity of the indirect method.

\section{Conclusion}\label{Sec:Conclusion}
In classical physics, the theory of synchronization has been extended from single particle synchronization into the synchronization between multi-particles with different properties. This inspire the present study of indirect quantum synchronization.

Examining two coupled two-level system, we found that an effective limit cycle exist in the system without drive, and the synchronization is feasible. We calculate the Husimi $Q$ function and discuss the phase locking feature of the system, then we analyse the $S$ function defined in the earlier publication to measure the synchronization. We find that indirect synchronization is possible. This result provide us with a method to synchronize a quantum system couples to its neighbour without adding the drive directly. The synchronization of multiple TLS, synchronization of two many-level systems is worth studying, this is beyond the scope of this paper and will be the topics of further studies.

In classical physics, the theory of synchronization has been extended from single particle synchronization into the synchronization between multi-particles with different properties. This inspire the present study of indirect quantum synchronization.

Examining two coupled two-level system, we found that an effective limit cycle exist in the system without drive, and the synchronization is feasible. We calculate the Husimi $Q$ function and discuss the phase locking feature of the system, then we analyse the $S$ function defined in the earlier publication to measure the synchronization. We find that indirect synchronization is possible. This result provide us with a method to synchronize a quantum system couples to its neighbour without adding the drive directly. The synchronization of multiple TLS, synchronization of two many-level systems is worth studying, this is beyond the scope of this paper and will be the topics of further studies.

\bibliographystyle{plain}

\begin{thebibliography}{9}

\bibitem{H. Fujisaka and T. Yamada(1983)}
T. Yamada and H. Fujisaka, Stability Theory of Synchronized Motion in Coupled-Oscillator Systems. II: The Mapping Approach.  \href{https://doi.org/10.1143/PTP.70.1240.}{Progress of Theoretical Physics. \textbf{70} (1983) 1240–1248.}
\bibitem{L. M. Pecora and T. L. Carroll(1990)}L.M. Pecora and T.L. Carroll. Synchronization in chaotic systems.  \href{https://doi.org/10.1103/PhysRevLett.64.821.}{Phys. Rev. Lett. \textbf{64} (1990) 821–824.}
\bibitem{M. Barahona and L. M. Pecora(2002)}[M. Barahona and L.M. Pecora. Synchronization in Small-World Systems.  \href{https://doi.org/10.1103/PhysRevLett.89.054101.}{Phys. Rev. Lett. \textbf{89} (2002) 054101.}
\bibitem{C Huygens}C. Huygens, Oeuvres Complete de Christiaan Huygens: Correspondence (Societe Hollandaise des Sciences, The Hague, The Netherlands, 1893).
\bibitem{I. Goychuk}I. Goychuk, J. Casado-Pascual, M. Morillo, J. Lehmann and P. Hänggi. Quantum Stochastic Synchronization. \href{https://doi.org/10.1103/PhysRevLett.97.210601.}{Phys. Rev. Lett. \textbf{97} (2006) 210601.}
\bibitem{M. Ludwig}M. Ludwig and F. Marquardt. Quantum Many-Body Dynamics in Optomechanical Arrays.  \href{https://doi.org/10.1103/PhysRevLett.111.073603.}{Phys. Rev. Lett. \textbf{111} (2013) 073603.}
\bibitem{A. Mari}A. Mari, A. Farace, N. Didier, V. Giovannetti and R. Fazio. Measures of Quantum Synchronization in Continuous Variable Systems.  \href{https://doi.org/10.1103/PhysRevLett.111.103605.}{Phys. Rev. Lett. \textbf{111} (2013) 103605.}
\bibitem{T. E. Lee}T.E. Lee and M.C. Cross. Quantum-classical transition of correlations of two coupled cavities.  \href{https://doi.org/10.1103/PhysRevA.88.013834.}{Phys. Rev. A \textbf{88} (2013) 013834.}
\bibitem{A. M. Hriscu}A.M. Hriscu and Yu.V. Nazarov. Quantum Synchronization of Conjugated Variables in a Superconducting Device Leads to the Fundamental Resistance Quantization.  \href{https://doi.org/10.1103/PhysRevLett.110.097002.}{Phys. Rev. Lett. \textbf{110} (2013) 097002.}

\bibitem{G. Manzano}G. Manzano, F. Galve, G.L. Giorgi, E. Hernández-García and R. Zambrini. Synchronization, quantum correlations and entanglement in oscillator networks.  \href{https://doi.org/10.1038/srep01439.}{Sci Rep. \textbf{3} (2013) 1439.}
\bibitem{S. E. Nigg}S.E. Nigg. Observing quantum synchronization blockade in circuit quantum electrodynamics.  \href{https://doi.org/10.1103/PhysRevA.97.013811.}{Phys. Rev. A \textbf{97} (2018) 013811.}
\bibitem{O. V. Zhirov and D. L. Shepelyansky(2009)}O.V. Zhirov and D.L. Shepelyansky. Quantum synchronization and entanglement of two qubits coupled to a driven dissipative resonator.  \href{https://doi.org/10.1103/PhysRevB.80.014519.}{Phys. Rev. B \textbf{80} (2009) 014519.}

\bibitem{V. Ameri(2015)}V. Ameri, M. Eghbali-Arani, A. Mari, A. Farace, F. Kheirandish, V. Giovannetti and R. Fazio. Mutual information as an order parameter for quantum synchronization.  \href{https://doi.org/10.1103/PhysRevA.91.012301.}{Phys. Rev. A \textbf{91} (2015) 012301.}

\bibitem{C. Davis}C. Davis-Tilley and A.D. Armour. Synchronization of micromasers.  \href{https://doi.org/10.1103/PhysRevA.94.063819.}{Phys. Rev. A \textbf{94} (2016) 063819.}

\bibitem{M. Xu(2014)}M. Xu, D.A. Tieri, E.C. Fine, J.K. Thompson and M.J. Holland. Synchronization of Two Ensembles of Atoms.  \href{https://doi.org/10.1103/PhysRevLett.113.154101.}{Phys. Rev. Lett. \textbf{113} (2014) 154101.}

\bibitem{M. Xu and M. J(2015)}M. Xu and M.J. Holland. Conditional Ramsey Spectroscopy with Synchronized Atoms.  \href{https://doi.org/10.1103/PhysRevLett.114.103601.}{Phys. Rev. Lett. \textbf{114} (2015) 103601.}

\bibitem{M. R. Hush(2015)}M.R. Hush, W. Li, S. Genway, I. Lesanovsky and A.D. Armour. Spin correlations as a probe of quantum synchronization in trapped-ion phonon lasers.  \href{https://doi.org/10.1103/PhysRevA.91.061401.}{Phys. Rev. A \textbf{91} (2015) 061401.}

\bibitem{N. Lorch(2017)}N. Lörch, S.E. Nigg, A. Nunnenkamp, R.P. Tiwari and C. Bruder. Quantum Synchronization Blockade: Energy Quantization Hinders Synchronization of Identical Oscillators.  \href{https://doi.org/10.1103/PhysRevLett.118.243602.}{Phys. Rev. Lett \textbf{118} (2017) 243602.}

\bibitem{T. E. Lee and H. R(2013)}T.E. Lee and H.R. Sadeghpour. Quantum Synchronization of Quantum van der Pol Oscillators with Trapped Ions.  \href{https://doi.org/10.1103/PhysRevLett.111.234101.}{Phys. Rev. Lett. \textbf{111} (2013) 234101.}

\bibitem{T. E. Lee(2014)}T.E. Lee, C.-K. Chan and S. Wang. Entanglement tongue and quantum synchronization of disordered oscillators.  \href{https://doi.org/10.1103/PhysRevE.89.022913.}{Phys. Rev. E \textbf{89} (2014) 022913.}

\bibitem{S. Walter(2014)}S. Walter, A. Nunnenkamp and C. Bruder. Quantum Synchronization of a Driven Self-Sustained Oscillator.  \href{https://doi.org/10.1103/PhysRevLett.112.094102.}{Phys. Rev. Lett. \textbf{112} (2014) 094102.}

\bibitem{M. Samoylova(2015)}M. Samoylova, N. Piovella, G.R.M. Robb, R. Bachelard and Ph.W. Courteille. Synchronization of Bloch oscillations by a ring cavity.  \href{https://doi.org/10.1364/OE.23.014823.}{Opt. Express \textbf{23} (2015) 14823.}

\bibitem{Y. Gul(2016)}Y. Gül, Synchronization of networked Jahn–Teller systems in SQUIDs.  \ref{https://doi.org/10.1142/S0217979216501253.}{Int. J. Mod. Phys. B \textbf{30} (2016) 1650125.}

\bibitem{Fernando}F. Quijandría, D. Porras, J.J. García-Ripoll and D. Zueco. Circuit QED Bright Source for Chiral Entangled Light Based on Dissipation.  \ref{https://doi.org/10.1103/PhysRevLett.111.073602.}{Phys. Rev. Lett. \textbf{111} (2013) 073602.}

\bibitem{Alexandre Roulet(1)}A. Roulet and C. Bruder. Synchronizing the Smallest Possible System.  \href{https://doi.org/10.1103/PhysRevLett.121.053601.}{Phys. Rev. Lett. \textbf{121} (2018) 053601.}


\bibitem{Alvaro Parra-Lopez(2020)}Á. Parra-López and J. Bergli. Synchronization in two-level quantum systems.  \href{https://doi.org/10.1103/PhysRevA.101.062104.}{Phys. Rev. A \textbf{101} (2020) 062104.}

\bibitem{Alexandre Roulet(2)}A. Roulet and C. Bruder. Quantum synchronization and entanglement generation.  \href{https://doi.org/10.1103/PhysRevLett.121.063601.}{Phys. Rev. Lett. \textbf{121} (2018) 063601.}

\bibitem{G. M}G.M. Huang, T.J. Tarn and J.W. Clark. On the controllability of quantum mechanical systems. \href{https://doi.org/10.1063/1.525634.}{Journal of Mathematical Physics. \textbf{24} (1983) 2608–2618.}

\bibitem{R. S}R.S. Judson and H. Rabitz. Teaching lasers to control molecules.  \href{https://doi.org/10.1103/PhysRevLett.68.1500.}{Phys. Rev. Lett. \textbf{68} (1992) 1500–1503.}

\bibitem{H. M}H.M. Wiseman. Quantum theory of continuous feedback.  \href{https://doi.org/10.1103/PhysRevA.49.2133.}{Phys. Rev. A \textbf{49} (1994) 2133–2150.}

\bibitem{A. C1}A.C. Doherty, S.M. Tan, A.S. Parkins and D.F. Walls. State determination in continuous measurement.  \href{https://doi.org/10.1103/PhysRevA.60.2380.}{Phys. Rev. A \textbf{60} (1999) 2380–2392.}

\bibitem{A. C2}A.C. Doherty and K. Jacobs. Feedback control of quantum systems using continuous state estimation.  \href{https://doi.org/10.1103/PhysRevA.60.2700.}{Phys. Rev. A \textbf{60} (1999) 2700–2711.}

\bibitem{A. C3}A.C. Doherty, S. Habib, K. Jacobs, H. Mabuchi and S.M. Tan. Quantum feedback control and classical control theory. \href{https://doi.org/10.1103/PhysRevA.62.012105.}{Phys. Rev. A \textbf{62} (2000) 012105.}

\bibitem{P. H}P.H. Bucksbaum. Particles driven to diffraction.  \href{https://doi.org/10.1038/35093182.}{Nature. \textbf{413} (2001) 117–118.}

\bibitem{S. Lloyd}S. Lloyd, A.J. Landahl and J.-J.E. Slotine. Universal quantum interfaces.  \href{https://doi.org/10.1103/PhysRevA.69.012305.}{Phys. Rev. A \textbf{69} (2004) 012305.}

\bibitem{D. Burgarth}D. Burgarth, K. Maruyama, M. Murphy, S. Montangero, T. Calarco, F. Nori and M.B. Plenio. Scalable quantum computation via local control of only two qubits.  \href{https://doi.org/10.1103/PhysRevA.81.040303.}{Phys. Rev. A \textbf{81} (2010) 040303.}

\bibitem{A. Kay}A. Kay and P.J. Pemberton-Ross. Computation on spin chains with limited access.  \href{https://doi.org/10.1103/PhysRevA.81.010301.}{Phys. Rev. A \textbf{81} (2010) 010301.}

\bibitem{A. C3}D. Burgarth and K. Yuasa. Quantum System Identification.  \href{https://doi.org/10.1103/PhysRevLett.108.080502.}{Phys. Rev. Lett. \textbf{108} (2012) 080502.}

\bibitem{M. Owari}M. Owari, K. Maruyama, T. Takui and G. Kato. Probing an untouchable environment for its identification and control.  \href{https://doi.org/10.1103/PhysRevA.91.012343.}{Phys. Rev. A \textbf{91} (2015) 012343.}

\bibitem{Kato}Kato, Owari, Maruyama. Hilbert Space Structure Induced by Quantum Probes. \href{https://doi.org/10.3390/proceedings2019012004.}{Proceedings. \textbf{12} (2019) 4.}

\bibitem{Michel Le Bellac(2006)}M. L. Bellac and P. Forcrand-Millard. Quantum Physics (Cambridge University Press, England, 2006).
\bibitem{H.P. Breuer}H. P. Breuer and F. Petruccione. The Theory of Open Quantum Systems (Oxford University Press, Oxford, 2007).
\bibitem{R. Gilmore(2018)}R. Gilmore, C.M. Bowden and L.M. Narducci. Classical-quantum correspondence for multilevel systems.  \href{https://doi.org/10.1103/PhysRevA.12.1019.}{Phys. Rev. A \textbf{12} (1975) 1019–1031.}

\bibitem{Arecchi F T}F.T. Arecchi, E. Courtens, R. Gilmore and H. Thomas. Atomic Coherent States in Quantum Optics.  \href{https://doi.org/10.1103/PhysRevA.6.2211.}{Phys. Rev. A \textbf{6} (1972) 2211–2237.}

\bibitem{A. Pikovsky}A. Pikovsky, M. Rosenblum and J. Kurths. Synchronization: A Universal Concept in Nonlinear Sciences, Cambridge Nonlinear Science Series (Cambridge University Press, Cambridge, England, 2001).
\bibitem{S. Sonar}S. Sonar, M. Hajdušek, M. Mukherjee, R. Fazio, V. Vedral, S. Vinjanampathy and L.-C. Kwek.  \href{https://doi.org/10.1103/PhysRevLett.120.163601.}{Squeezing Enhances Quantum Synchronization, Phys. Rev. Lett. \textbf{120} (2018) 163601.}
\bibitem{H. J. Carmichael}H. J. Carmichael. An Open System (Springer-Verlag, Berlin, 1999).
\bibitem{E. H}H. Eneriz, D.Z. Rossatto, F.A. Cárdenas-López, E. Solano and M. Sanz. Degree of Quantumness in Quantum Synchronization.  \href{https://doi.org/10.1038/s41598-019-56468-x.}{Sci Rep. \textbf{9} (2019) 19933.}

\bibitem{Koppenhofer}M. Koppenhöfer and A. Roulet. Optimal synchronization deep in the quantum regime: Resource and fundamental limit.  \href{https://doi.org/10.1103/PhysRevA.99.043804.}{Phys. Rev. A \textbf{99} (2019) 043804.}
\end{thebibliography}

\end{document}